\documentclass[aps,pra,twocolumn,reprint]{revtex4-1}
\usepackage{blindtext}

\usepackage{graphicx}
\usepackage{amssymb}
\usepackage{amsmath}
\usepackage{bm}
\usepackage{gensymb}
\usepackage{footmisc}

\begin{document}
\title{Double resonance of Raman transitions in a degenerate Fermi gas}
\author{Moosong Lee, Jeong Ho Han, Jin Hyoun Kang, Min-Seok Kim, Y. Shin}
\email{yishin@snu.ac.kr}

\affiliation{
Center for Correlated Electron Systems, Institute for Basic Science, Seoul 08826, Korea
}
\affiliation{
Department of Physics and Astronomy and Institute of Applied Physics, Seoul National University, Seoul 08826, Korea
}

\date{\today}

\begin{abstract}
We measure momentum-resolved Raman spectra of a spin-polarized degenerate Fermi gas of $^{173}$Yb atoms for a wide range of magnetic fields, where the atoms are irradiated by a pair of counterpropagating Raman laser beams as in the  conventional spin-orbit coupling scheme. Double resonance of first- and second-order Raman transitions occurs at a certain magnetic field and the spectrum exhibits a doublet splitting for high laser intensities. The measured spectral splitting is quantitatively accounted for by the Autler--Townes effect. We show that our measurement results are consistent with the spinful band structure of a Fermi gas in the spatially oscillating effective magnetic field generated by the Raman laser fields.  

\end{abstract}

\maketitle

\section{Introduction}

Spin-orbit coupling (SOC) interwines the motional degrees of freedom of a system with its spin part, giving rise to many intriguing phenomena such as the atomic fine structure, the spin Hall effect~\cite{Kato2004, Konig2007} and topological insulators~\cite{Hasan2010}. In ultracold atom experiments, SOC has been realized using Raman laser dressing techniques~\cite{Lin2011, Goldman2013e, Zhai2015a}, where a two-photon Raman transition couples two different spin-momentum states. This optical method was successfully applied to many fermionic atom systems~\cite{Wang2012, Cheuk2012b, Song2016, Burdick2016} and recently extended to two-dimensions~\cite{Huang2016}, boosting the interest in exploring new exotic SOC-driven many-body phenomena~\cite{Zhai2015a,Cao2014, Xu2014}. 

Alkaline-earth-like atoms with two valence electrons such as ytterbium and strontium provide a beneficial setting for studying SOC physics. Their transition linewidth is narrow in comparison to the hyperfine structure splitting, which is helpful to alleviate the unavoidable heating effect due to light-induced spontaneous scattering under the Raman dressing~\cite{Zhai2015a, Song2016} and also to generate spin-dependent optical coupling to the hyperfine ground states. Furthermore, as recently demonstrated with $^{173}$Yb atoms~\cite{Pagano2015,Hofer2015}, the interorbital interactions between the $^1S_0$ and $^3P_0$ states can be tuned via a so-called orbital Feshbach resonance~\cite{Zhang15}, which would broaden the research scope of the SOC physics with alkaline-earth-like atoms.

In this paper, we present momentum-resolved Raman spectra of a spin-polarized degenerate Fermi gas of $^{173}$Yb atoms, which are measured in the Raman laser configuration of the conventional SOC scheme. In particular, we measure the Raman spectra over a wide range of magnetic fields as well as laser intensities to investigate the interplay of multiple Raman transitions in the SOC scheme. We observe that two Raman transitions become simultaneously resonant at a certain magnetic field and a doublet structure develops in the spectrum for strong Raman laser intensities. We find that the spectral splitting at the double resonance is quantitatively accounted for by the Autler--Townes doublet effect~\cite{Autler1955}. 

In the conventional SOC scheme, since one of the Raman laser beams has both $\sigma^+$ and $\sigma^-$ polarization components with respect to the quantization axis defined by the magnetic field, the Raman transition from one spin state to another, if any, can be made to impart momentum in either direction along the relative Raman beam propagation axis. In typical SOC experiments, the system parameters are set to make one of the transitions energetically unfavorable such that it can be ignored, but the double resonance observed in this work results from involving both of the Raman transitions. When all the Raman transitions are taken into account, the effect of the Raman laser fields is represented by a spatially oscillating effective magnetic field~\cite{Jimenez-Garcia2012}. We show that our measurement results are consistent with the spinful band structure of the Fermi gas under the effective magnetic field.

The paper is organized as follows. In Sec.~II, we describe our experimental apparatus and procedures for sample preparation and Raman spectroscopy. In Sec.~III, we present the Raman spectra measured for various conditions and the observation of the spectral doublet splitting at the double resonance. In Sec.~IV, we discuss the results in the perspective of the spinful band structure of the SO-coupled Fermi gas. Finally, a summary and outlooks are provided in Sec. V.

\section{Experiments}

Figure~1(a) shows the schematic diagram of our experimental apparatus for generating a degenerate Fermi gas of $^{173}$Yb atoms~\cite{Yb_machine}. We first collect ytterbium atoms with a Zeeman slower and a magneto optical trap (MOT). For the slowing light, we use a 399~nm laser beam that has a dark spot at its center to suppress the detrimental scattering effect on atoms in the MOT. The frequency modulation method is adopted for the 556~nm MOT beams to increase the trapping volume and capture velocity of the MOT~\cite{Fukuhara2007a}. As a result, more than $10^8$ atoms are collected in the MOT within 15~s. We transfer the atoms into an optical dipole trap (ODT) formed by a focused 1070~nm laser beam, where the transfer efficiency is $\approx 13\%$. Then, we transport the atoms by moving the ODT to a small appendant chamber which provides better optical access and allows high magnetic field application, and we generate a crossed ODT by superposing a focused 532~nm laser beam horizontally with the 1070~nm ODT. 

After evaporation cooling, we obtain a quantum degenerate sample in the $F=5/2$ hyperfine ground state. For an equal mixture of the six spin components, the total atom number is $N\approx 1.0\times 10^5$ and the temperature is $T/T_F \approx 0.1$, where $T_F$ is the Fermi temperature of the trapped sample. The spin composition of the sample can be manipulated during evaporative cooling with optical pumping or removal of spin states by resonant light. For the case of a fully spin-polarized sample in the $m_F=-5/2$ state, $N\approx 1.2\times10^5$ and $T/T_F \approx 0.35$. The trapping frequencies of the crossed ODT are $(\omega_r, \omega_z) = 2\pi\times(52, 450)$~Hz at the end of the sample preparation.

\begin{figure}
\includegraphics[width=8.2cm]{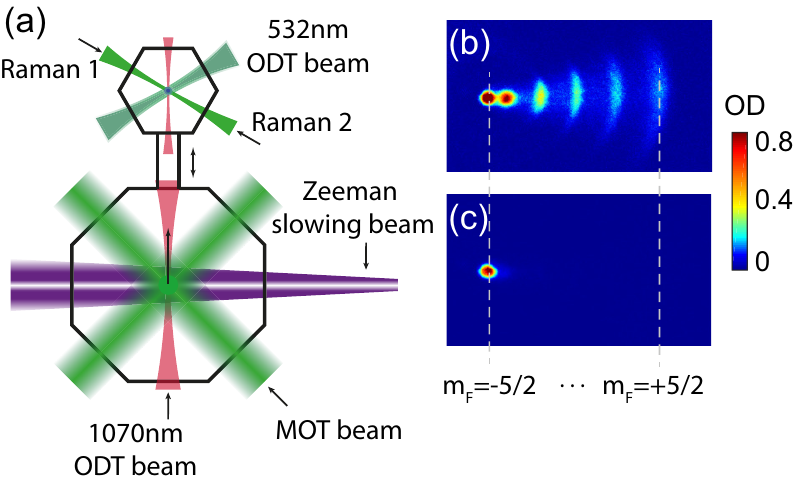}
\caption{(a) Schematic diagram of the experimental apparatus. $^{173}$Yb atoms are collected with a Zeeman slower and a magneto-optical trap (MOT), loaded into a 1070~nm optical dipole trap (ODT), and transported to a small hexagonal chamber by moving the ODT. (b, c) Images of atoms after optical Stern-Gerlach spin separation~\cite{Taie2010, Stellmer2011}: equal mixture of the six spin components of the $F=5/2$ hyperfine ground state (b) and fully polarized sample in the $m_F=-5/2$ spin state (c).
}
\label{fig:sample_preparation}
\end{figure}

The setup for Raman spectroscopy is illustrated in Fig.~2(a). A pair of counter-propagating laser beams are irradiated on the sample in the $x$ direction and an external magnetic field $B$ is applied in the $z$ direction. The two laser beams are linearly polarized in the $y$ and $z$ directions, respectively. With respect to the quantization axis defined by the magnetic field in the $z$ direction, Raman beam 1 with linear $y$  polarization has both $\sigma^+$ and $\sigma^-$ components and Raman beam 2 with linear $z$ polarization has a $\pi$ component. Thus, a two-photon Raman process, e.g., imparting momentum of $+2\hbar k_R \hat{x}$ by absorbing a photon from Raman beam 1 and emitting a photon into Raman beam 2 changes the spin number by either $+1$ or $-1$, where $k_R$ is the wavenumber of the Raman beams. This is the conventional  Raman laser configuration for SOC in cold atom experiments~\cite{Lin2011, Wang2012, Cheuk2012b, Burdick2016}. 

The Raman lasers are blue-detuned by 1.97~GHz from the $|^1S_0, F = 5/2\rangle$ to $|^3P_1, F'=7/2\rangle$ transition~[Fig.~2(b)]. This laser detuning, set between the hyperfine states of $^3P_1$, is beneficial to induce spin-dependent transition strengths for the $F=5/2$ hyperfine spin states~\cite{Mancini2015, Mancini2015a}. The frequency difference of the two Raman beams is denoted by $\delta \omega$ [Fig.~2(a)]. The two beams are set to the same power $P$ and focused onto the sample. Their $1/e^2$ intensity radii are $\approx 150~\mu$m, which is much larger than the trapped sample size of 30~$\mu$m. We assume that the laser intensities are uniform over the sample.

\begin{figure}
\includegraphics[width=8.2cm]{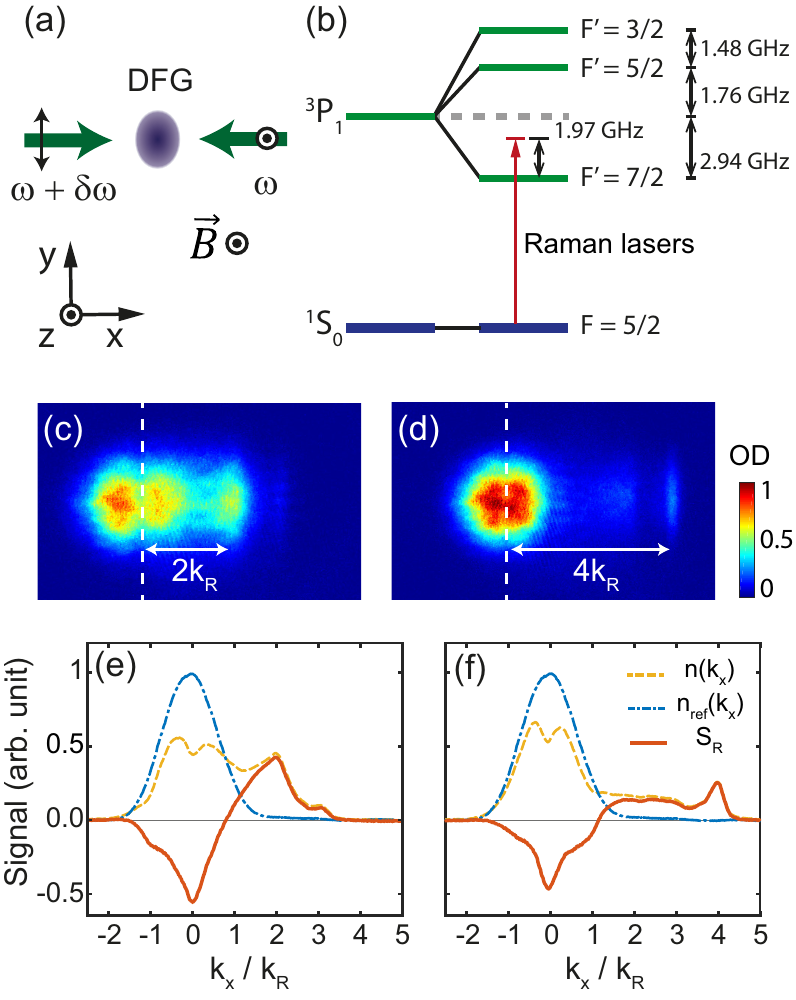}
\caption{Raman spectroscopy of a degenerate Fermi gas (DFG). (a) Raman coupling setup with a pair of counter-propagating laser beams, whose frequency difference is denoted by $\delta\omega$. (b) Energy diagram of the $^3P_1$ state of $^{173}$Yb and the relative detuning of the Raman laser. (c, d) Examplary time-of-flight images of Fermi gases after applying a pulse of the Raman beams for $\delta \omega/2\pi = 14.8$~kHz (c) and $29.6$~kHz (d). The vertical dashed lines indicate the center of the unperturbed sample. (e, f) 1D momentum distributions $n(k_x)$ of the samples (yellow dashed curve) obtained by integrating the images along the $y$ direction. The normalized Raman spectra $S_R(k_x)$ (red solid curve) are measured as $S_R(k_x)=[n(k_x)-n_\textrm{ref}(k_x)]/n_\textrm{ref}(0)$, where $n_\textrm{ref}(k_x)$ is the reference distribution (blue dash-dotted curve) obtained without applying the Raman beam pulse. 
}
\label{fig:raman_setup}
\end{figure}

Raman spectroscopy is performed by applying a pulse of the Raman beams and taking a time-of-flight absorption image of the sample. The image is taken at $B=0$~G along the $z$-axis with a linearly polarized probe beam resonant to the $^1S_0\rightarrow {^1}P_1$ transition. Two exemplary images are shown in Fig.~2(c) and 2(d), showing that atoms are scattered out of the original sample with different momenta for different $\delta\omega$. Since the expansion time $\tau$ is sufficiently long such that $\omega_r \tau \approx 5$, we interpret the time-of-flight image as the momentum distribution of the atoms. The 1D momentum distribution $n(k_x)$ is obtained by integrating the image along the $y$ direction [Fig.~2(e) and 2(f)], where $k_x=mx/(\hbar \tau)$ with $m$ being the atomic mass and $x$ the displacement from the center of mass of an unperturbed sample. In our imaging, the absorption coefficient for each spin state was found to vary slightly, within $\approx 10$\%~[Fig.~1(b)], which we ignored in the determination of $n(k_x)$. 

The normalized Raman spectrum is measured as  
\begin{equation}
S_R(k_x) = [n(k_x) - n_\mathrm{ref}(k_x)]/n_\mathrm{ref}(0),
\end{equation}
where $n_\mathrm{ref}$ is the reference distribution obtained without applying the Raman beams. In the spectrum, a momentum-imparting Raman transition appears as a pair of dip and peak, which correspond to the initial and final momenta of the transition, respectively. We observe that the spectral peaks and dips exhibit slightly asymmetric shapes, which we attribute to elastic collisions of atoms during the time-of-flight expansion~\cite{Veeravalli2008}. The Fermi momentum of the sample is $k_F/k_R\approx 1.2$ in units of the recoil momentum.

\section{Results}

The atomic state in an ideal Fermi gas is specified by wavenumber $k$ and spin number $m_F$, and its energy level is given by   
\begin{equation} \label{eq:energy} 
E(|k,m_F\rangle) = \frac{\hbar^2 k^2}{2m}  + g_F \mu_B m_F B + E_S(m_F).
\end{equation}
The first term is the kinetic energy of the atom and the second term is the Zeeman energy due to the external magnetic field $B$, where $g_F$ is the Land\'{e} $g$-factor and $\mu_B$ is the Bohr magneton. The last term $E_S$ denotes the spin-dependent ac Stark shift induced by the Raman lasers. For a Raman transition from $|k_i, m_i\rangle$ to $|k_f=k_i+2r k_R, m_f=m_i+\Delta m_F\rangle$, which changes the momentum by $2 r \hbar k_R$ and the spin number by $\Delta m_F$, the energy conservation requires $E(|k_f,m_f\rangle)-E(|k_i,m_i\rangle)=r\hbar \delta \omega$, which gives the resonance condition for the initial wavenumber $k_i$ as
\begin{equation} \label{eq:resonance} 
k_i = k_R \Big[ \frac{\hbar \delta\omega}{4E_R} - \frac{\Delta m_F}{r} \frac{B}{4B_R}-\frac{1}{r}\frac{\Delta E_S}{4E_R} - r \Big],
\end{equation}
where $E_R = (\hbar k_R)^2/2m = h \times 3.7$~kHz is the atomic recoil energy, $B_R=E_R/(g_F \mu_B)=17.9$~G and $\Delta E_S=E_S(m_f)-E_S(m_i)$. Here we neglect the quadractic Zeeman effect and the atomic interactions which are negligible in our experimental conditions.

\begin{figure}
\includegraphics[width=8.4cm]{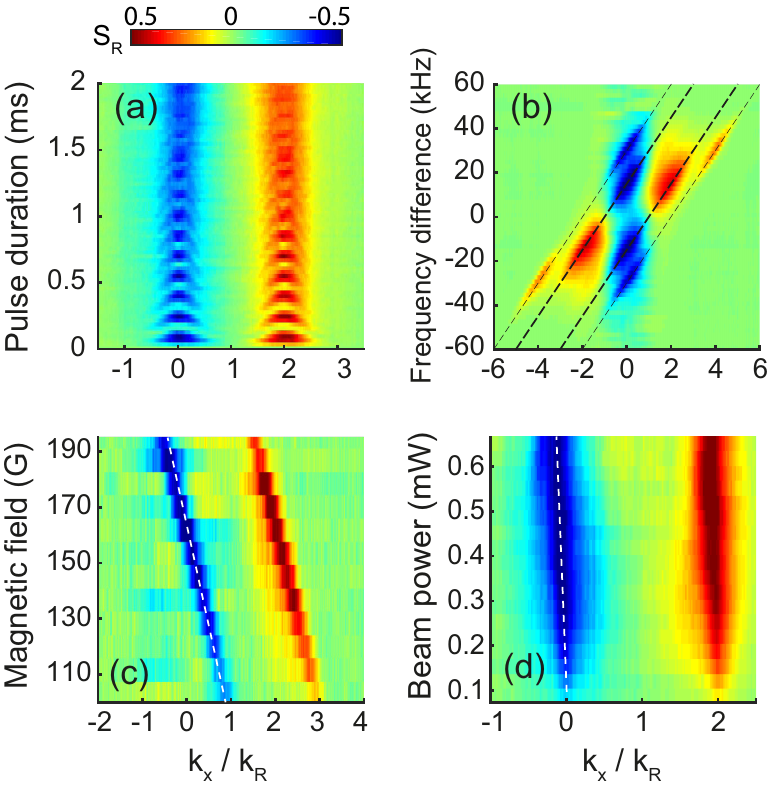}
\caption{Raman spectra measured by scanning various experimental parameters including Raman beam pulse duration $t$, frequency difference $\delta \omega$, magnetic field $B$, and Raman beam power $P$: (a) $\delta\omega = 4E_R/\hbar$, $B=~16.6$~G, $P=0.47$~mW; (b) $t=2$~ms, $B=0$~G, $P=1.1$~mW; (c) $t=2$~ms, $\delta \omega= 13.4 E_R/\hbar$, $P=0.21$~mW; (d) $t=2$~ms, $\delta \omega = 13.4 E_R/\hbar$, $B=133$~G (see the text for details of the sample condition and the polarization configuration of the Raman beams). The dashed lines in (b) indicate $k=k_R(\frac{\hbar}{4E_R}\delta \omega -n)$ for $n=-2,-1,1$,and 2, and those in (c) and (d) are guides for the eyes having slopes of $\frac{dk}{d B}=-\frac{k_R}{4B_R}$ and $\frac{dk}{d P}=-0.3 k_R/$mW. Each spectrum was obtained by averaging more than three measurements.
}
\label{fig:Raman_Rabi}
\end{figure}

\begin{figure*}
\includegraphics[width=17.0cm]{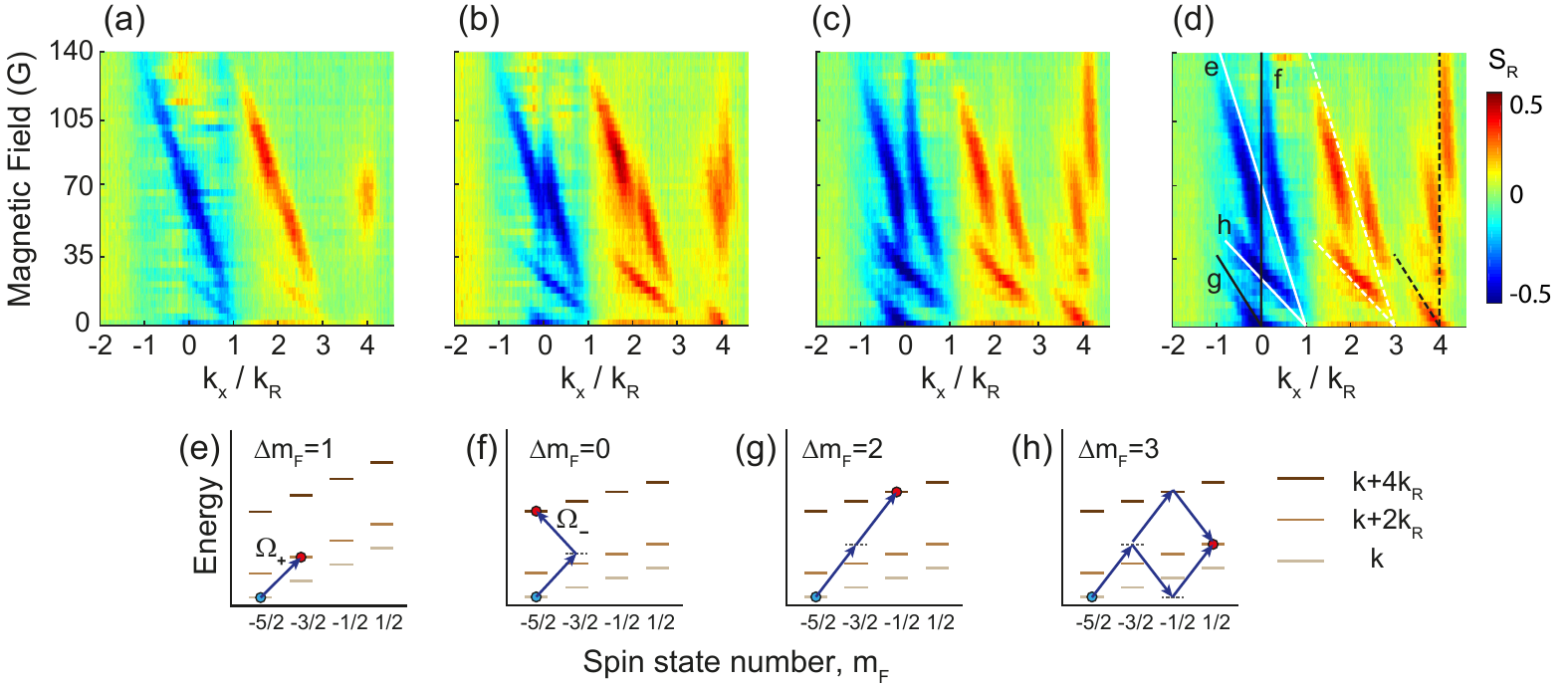}
\caption{Double resonance of Raman transitions. (a--c) Raman spectra of an $m_F=-5/2$ spin-polarized sample as a function of the magnetic field $B$ for $\delta \omega =8 E_R/\hbar$ and various Raman beam powers (a) $P = 0.13$~mW, (b) 0.21~mW, and (c) 0.36~mW. As the Raman coupling strength increases with higher $P$, a spectral doublet splitting develops at $B=4 B_R\approx 72$~G where the $(r,\Delta m_F)=(1, 1)$ and $(2, 0)$ transitions are doubly resonant. The spectrum in (d) is the same of (c) with the guide lines (solid) indicating the resonant momentum positions for various Raman transitions, which are calculated from Eq.~(3) without including the ac Stark shift. The dashed lines are the corresponding final momentum positions. (e--h) Diagrams of the Raman transitions observed in the spectra.   
}
\label{fig:highorder}
\end{figure*}

We first investigate the resonance condition of Eq.~(3) by measuring its dependence on various experimental parameters. Figure 3(a) shows a Raman spectrum measured by scanning the Raman beam pulse duration for $\delta \omega =4 E_R/\hbar$ at $B=16.6$~G. Spin-polarized samples were used and both of the Raman beams were set to linear $z$ polarization to make sure $\Delta m_F=0$.  Momentum-dependent Rabi oscillations are clearly observed and the Rabi frequency is found to be well described by $\Omega(k) = \sqrt {\Omega_0^2 + (\hbar k_R k /m)^2 }$ with $\Omega_0 \approx 2\pi\times7$~kHz. The decoherence time is measured to be $\approx 1$~ms, which seems to be understandable with the characteristic time scale for momentum dephasing in the trap, $\pi/(2\omega_r) \approx 5$~ms. In the following, we set the pulse duration of the Raman beam to 2~ms, which is long enough to study the steady state of the system under the Raman laser dressing. 

Figure~3(b) displays a spectrum of the equal mixture sample in the plane of wavenumber $k$ and frequency difference $\delta \omega$. Here, $B=0$~G and the Zeeman effect is absent in the measurement. The $r=1$ and $r=2$ transitions are identified in the spectrum with their spectral slope of $\frac{d k}{d \delta\omega}= \frac{\hbar k_R}{4E_R}$ and different offsets as predicted by Eq.~(3). The $(k,\delta \omega)\leftrightarrow (-k,-\delta \omega)$ symmetry of the spectrum indicates that the differential ac Stark shift is negligible in the measurement.

Figure~3(c) shows the Raman spectrum of the $m_F=-5/2$ spin-polarized sample over a range of magnetic fields from $B=100$~G to 195~G for $\delta \omega = 13.4 E_R/\hbar$. In the spectral plane of $k$ and $B$, the Raman transition with $(r,\Delta m_F) = (1, 1)$ appears as a line having the slope $\frac{dk}{dB}=-\frac{k_R}{4 B_R}$ as expected from Eq.~(3). A linear spectral shift is observed with increasing Raman beam power $P$~[Fig.~3(d)], which demonstrates the effect of the differential ac Stark shift $\Delta E_S$. In our experiment, $\Delta E_S = E_S(-3/2)-E_S(-5/2)\approx 1.2~E_R$ for $P = 1$~mW. This is in good a agreement with the Raman beam intensities estimated from the Rabi oscillation frequency $\Omega_+ \propto \sqrt{I_\sigma I_{\pi}}$, where $I_{\sigma,\pi}$ are the intensities of the Raman beam 1 and 2, respectively. The comparison of $\Delta E_s$ and $\Omega_+$ suggests $I_{\pi} = 0.6I_{\sigma}$, which we attribute to a slight mismatch of the beam waists.

Next we investigate a situation where one spin-momentum state is resonantly coupled to two final states simultaneously, which we refer to as a double resonance. When the two corresponding Raman processes are characterized with $(r_1, \Delta m_{F1})$ and $(r_2, \Delta m_{F2})$, we see from Eq.~(\ref{eq:resonance}), neglecting the small $\Delta E_S$ term, that the double resonance occurs when 
\begin{equation} \label{eq:double} 
\frac{B}{4 B_R} \frac{\Delta m_{F1}}{r_1} + r_1 = \frac{B}{4 B_R} \frac{\Delta m_{F2}}{r_2} + r_2.
\end{equation}
For the primary transition with $(r_1,\Delta m_{F1})=(1,1)$, the double resonance condition is satisfied at $B= \frac{4 r_2(r_2-1)}{r_2-\Delta m_{F2}} B_R$. 

To observe the double resonance of the $(r,\Delta m_F)=(1,1)$ and $(2,0)$ transitions at $B=4B_R\approx 72$~G, we measure the Raman spectra of the spin-polarized sample in the $k$-$B$ plane over a range from $B=0$~G to 140~G [Fig.~4]. Here we set $\delta \omega =8 E_R /\hbar$ to have $k_x=0$ atoms on resonance for the $(2,0)$ transition, which is insensitive to $B$ for $\Delta m_F=0$. For low $P$, the $(1,1)$ transition appears with the spectral slope of $-\frac{k_R}{4B_R}$ as observed in Fig.~3(c) and the double resonance is indicated by a small signal at $(k,B)=(4 k_R, 4 B_R)$~[Fig.~4(a)]. This is understood as enhancement of the second-order Raman transition from $|k=0,-5/2\rangle$ to $|k=4 k_R, -5/2\rangle$ due to its intermediate state $|k=2k_R, -3/2\rangle$ being resonant. When the Raman beam power increases, we observe development of a spectral splitting at the resonance~[Figs.~4(b) and 4(c)]. The overall pattern of the high-$P$ spectrum shows the avoided crossing of the spectral lines corresponding to the two $(1,1)$ and $(2,0)$ transitions. 

Near the double resonance, the system can be considered as a three-level system consisting of $|k,-5/2\rangle$, $|k+2k_R, -3/2\rangle$ and $|k+4k_R, -5/2\rangle$~[Fig.~5(a)]. For simplicity, we denote them by $|0\rangle$, $|1\rangle$, and $|2\rangle$, respectively. Since the Raman transition between $|0\rangle$ and $|1\rangle$ involves the $\sigma^+$ component of Raman beam 1 but that between $|1\rangle$ and $|2\rangle$ involves the $\sigma^-$ component, the coupling strengths $\Omega_+$ and $\Omega_-$ of the two transitions, respectively, can be different. In our case with $^{173}$Yb atoms in the $m_F=-5/2$ state, $\Omega_-=5.3~\Omega_+$. Since the coupling between $|1\rangle$ and $|2\rangle$ are much stronger than that between $|0\rangle$ and $|1\rangle$, the observed spectral splitting with increasing Raman beam intensity can be described as an Autler--Townes doublet~\cite{Autler1955}: two dressed states $|\alpha\rangle$ and $|\beta\rangle$ are formed with $|1\rangle$ and $|2\rangle$ under the strong coupling and their energy level splitting is probed via Raman transitions from the initial $|0\rangle$ state. In the rotating wave approximation, the energy levels of the two dressed states are given by $E_{\alpha, \beta} = \frac{1}{2}[E_1 + E_2 -3\hbar\delta\omega \pm \sqrt{(E_1 - E_2 + \hbar\delta\omega)^2 + (\hbar\Omega_-)^2}]$, where $E_{1,2}=E(|1,2\rangle)$. The resonant wavenumbers $k_{\alpha,\beta}$ of the initial state $|0\rangle$ are determined from $E(|0\rangle)=E_{\alpha,\beta}$ and for $\delta\omega=8E_R/\hbar$ and $B=4B_R$, we obtain $k_{\alpha,\beta}=\pm \frac{k_R}{8\sqrt{2}}\frac{\hbar \Omega_-}{E_R}$. We find our measurement results on the double resonance at $B=4B_R$ in good quantitative agreement with the estimation [Fig.~6(b)]. The coupling strength $\Omega_-$ was separately measured from the Rabi oscillation data of the $|0,-5/2\rangle \rightarrow |-2k_R,-3/2\rangle$ transition for $\delta \omega= -13.4 E_R/\hbar$ at $B=166$~G.

\begin{figure}
\includegraphics[width=8.0cm]{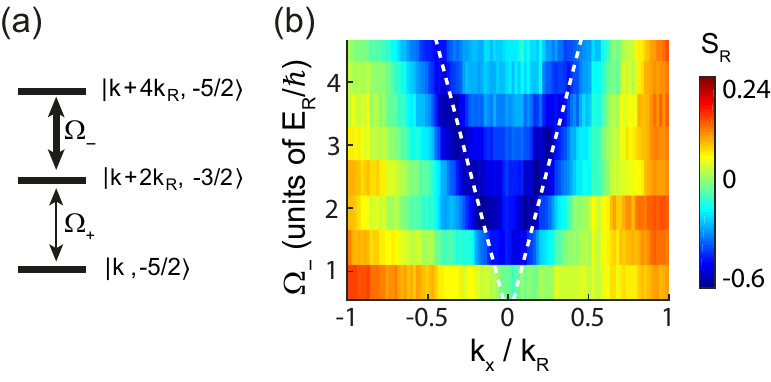}
\caption{Spectral splitting at double resonance. (a) Three atomic states involved in the double resonance at $B = 4B_R$. For a $^{173}$Yb atom in the $m_F=-5/2$ state, $\Omega_- = 5.3~\Omega_+$ and the upper two states are more strongly coupled.
(b) Raman spectrum for $\delta \omega = 8E_R/\hbar$ and $B= 4B_R$ as a function of the Raman coupling strength $\Omega_-$. The dashed lines are the theoretical prediction of $k_{\alpha,\beta}=\pm \frac{k_R}{8\sqrt{2}}\frac{\hbar \Omega_-}{E_R}$, which is calculated in the limit of $\Omega_+/\Omega_- \rightarrow 0$ (see text). 
}
\label{fig:gap}
\end{figure}

The Raman spectra in Fig.~4 reveal another double resonance at $B=\frac{4}{3} B_R\approx 24$~G, where the $(r,\Delta m_F)=(2,0)$ line crosses the $(r,\Delta m_F)=(1,3)$ line. Although the $(1,3)$ transition is a third-order Raman transition, its spectral strength is observed to be higher than that of the (2,0) transition. In the intermediate region of $B\approx 35$~G, many Raman transitions are involved over the whole momentum space of the sample and the spectral structure for high Raman laser intensity shows interesting features which cannot be simply explained as crossing and avoided crossing of the spectral lines. It might be necessary to take into account the ac Stark shift effect and a further quantitative analysis of the Raman spectra will be discussed in future work.

\begin{figure}
\includegraphics[width=8.0cm]{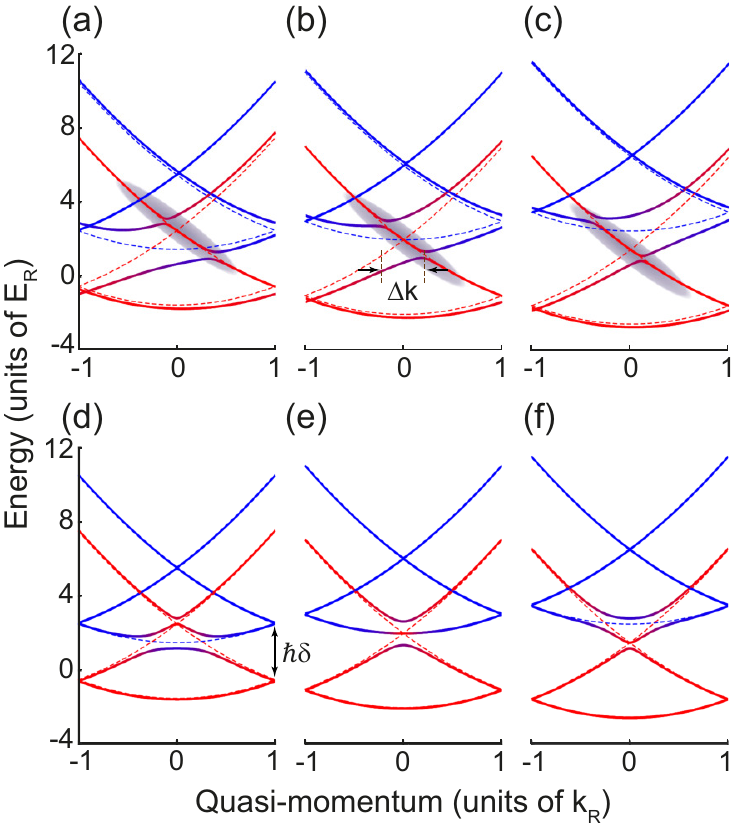}
\caption{Energy band structures of a SO-coupled spin-1/2 atom under the Raman laser dressing: $\Omega_-=5.3~\Omega_+$ in (a--c) and $\Omega_-=\Omega_+$ in (d--e), and $\hbar\delta = 3 E_R$ in (a, d), $4 E_R$ in (b, e), and $5 E_R$ in (c, f). The color of the solid lines indicates the bare spin fractions of the energy eigenstates: blue for spin-up and red for spin-down. The dashed lines represent the energy spectrum for $\Omega_\pm=0$. The gray shadow areas in (a--c) indicate the region corresponding to the momentum states occupied by the sample in the experiment. 
}
\label{fig:band}
\end{figure}

\section{Discussion}

In the Raman laser dressing scheme described in Fig.~2(a), two ways of couplings are allowed between the two spin states because the Raman beam that has linear polarization orthogonal to the magnetic field contains both $\sigma^+$ and $\sigma^-$ components. This means that a Raman transition from one spin state to the other spin state can occur while imparting momentum in either of the $x$ and $-x$ directions. In typical experimental conditions~\cite{Zhai2015a, Lin2011, Cheuk2012b, Wang2012}, one of the couplings is resonantly dominant over the other, giving rise to a form of SOC that has equal strengths of the Rashba and Dresselhaus contributions. However, when the Fermi sea of a sample covers a large momentum space, this approximation cannot be applied and it is necessary to include both of the Raman couplings for the full description of the system. Furthermore, as observed in the previous section, the two Raman couplings can be doubly resonant and play cooperative roles in the SOC physics of the system.

As an archetypal situation, we consider a spin-1/2 atom under the Raman dressing for $\delta \omega=0$. Here, the counterpropagating Raman beams form a stationary polarization lattice with spatial periodicity of $\pi/k_R$. Including all the allowed Raman transitions, the effective Hamiltonian of the system is given by
\begin{eqnarray}\label{eq:SOHamiltonian}
H&=&
  \begin{bmatrix}
    \frac{\hbar^2 k^2}{2m} + \frac{\hbar \delta}{2} & \frac{\hbar \Omega_+ }{2} e^{i2k_R x} + \frac{\hbar \Omega_- }{2} e^{-i2k_R x} \\
    \frac{\hbar \Omega_+}{2} e^{-i 2k_R x} + \frac{\hbar \Omega_-}{2} e^{i 2k_R x} & \frac{\hbar^2 k^2 }{2m} -\frac{\hbar \delta}{2} 
  \end{bmatrix} \nonumber \\
&=&
\frac{\hbar^2 k^2}{2m} + \frac{\hbar\delta}{2} \sigma_z \nonumber \\ 
&&~~+ \frac{\hbar}{2}\Omega_x\cos(2k_R x)\sigma_x + \frac{\hbar}{2}\Omega_y \sin(2k_R x)\sigma_y,
\end{eqnarray}
where $\hbar \delta$ is the sum of the differential Zeeman and ac Stark shifts, $\sigma_i$ are the $2\times2$ Pauli matrices, and $\Omega_{x,y} = \Omega_{+}\pm\Omega_{-}$. The final form of $H$ shows that the Raman dressing is equivalent to an effective magnetic field ${\cal{B}}=(\delta,~\Omega_x \cos (2k_R x),~\Omega_y \sin (2k_R x))$, which has two parts: a bias field along the $z$ axis and a spatially oscillating field on the $x$-$y$ plane. Its chirality is determined by the sign of $\Omega_x \Omega_y = \Omega_+^2 - \Omega_-^2$. In the presence of the spatially oscillating magnetic field, the energy dispersion of the atom has a spinful band structure [Fig.~6]. 

Figure 6(b) displays a band structure for $\Omega_-=5.3~\Omega_+$ and $\delta=4E_R/\hbar$, which straightforwardly explains the observed spectral splitting at the double resonance. In the experiment, $\delta \omega=8E_R/\hbar$ and the polarization lattice of the Raman beams moves in the lab frame with velocity of $+2\hbar k_R /m \hat{x}$. Initially, the atoms in the trapped sample occupy the low quasi-momentum region of the second and third bands of the bare spin-down state, which is indicated by a gray region in Fig.~6(b), and they are projected to the eigenstates of the spinful band structure via the Raman spectroscopy process. The quasi-momentum separation between the gap opening positions, which is marked with $\Delta k$ in Fig.~6(b), is the spectral splitting observed in our Raman spectrum. We note that $\Delta k=0$ in the symmetric case of $\Omega_-=\Omega_+$ [Fig.~6(e)] and the spectral splitting would not occur in the Raman spectrum.

\section{Summary and outlook}
We have measured the Raman spectra of a spin-polarized degenerate Fermi gas of $^{173}$Yb atoms in the conventional SOC scheme and investigated the double resonance of Raman transitions. We observed the development of a spectral splitting at the double resonance of the $(r,\Delta m_F)=(1,1)$ and $(2,0)$ transitions and provided its quantiative explanation as the Autler--Townes doublet effect. Finally, we discussed our results in the context of the spinful energy band structure under the Raman laser dressing. 

In general, when the system has multiple SOC paths in its spin-momentum space, a spinful energy band structure is formed because of the periodicity imposed by them. In previous experiments~\cite{Jimenez-Garcia2012, Cheuk2012b}, spinful band structures were designed and demonstrated by applying a RF field to the SO-coupled systems under the Raman laser dressing, where the role of the RF field was to open an additional coupling path between the two spin states. The results in this work highlight that the conventional Raman laser dressing scheme  provides two ways of SOC and intrinsically generates a spinful band structure without the aid of an additional RF field. An interesting extention of this work is to investigate the magnetic ordering and properties of a Fermi gas in the spatially rotating magnetic field ${\cal{B}}$. In the $F=5/2$ $^{173}$Yb system, the chirality of ${\cal{B}}$ can be controlled to some extent by the choice of the two spin states that are coupled by the Raman lasers. If the $m_F = \pm1/2$ states are employed, $\Omega_y=0$ and ${\cal{B}}$ changes from an axial field to a alternating transverse field as a function of $\hbar\delta$. In particular, when $\hbar\delta=0$, ${\cal{B}}=0$ points are periodically placed, which might profoundly affect the magnetic properties of the system. It was discussed in Ref.~\cite{Cheuk2012b} to engineer a flat spinful band structure, which might be pursued via proper tuning of the parameters of our system.

\section{Acknowledgments}
This work was supported by IBS-R009-D1 and the National Research Foundation of Korea (Grant No. 2014-H1A8A1021987).

\end{document}